\documentclass{optica-article}

\journal{optcon}

\articletype{Research Article}

\usepackage{graphicx}

\usepackage{lineno}

\begin{document}

\title{Spectroscopic localization of atomic sample plane for precise digital holography}

\author{Jian Zhao,\authormark{1} Yuzhuo Wang,\authormark{2} Xing Huang,\authormark{1} and Saijun Wu\authormark{1,*}}

\address{\authormark{1}Department of Physics, State Key Laboratory of Surface Physics and Key Laboratory of Micro and Nano Photonic Structures (Ministry of Education),Fudan University, Shanghai 200433, China.\\
\authormark{2}State Key Laboratory of Quantum Optics and Quantum Optics Devices, Institute of Laser Spectroscopy, Shanxi University, Taiyuan, Shanxi 030006, China.}
\email{\authormark{*}saijunwu@fudan.edu.cn}




\begin{abstract}
In digital holography, the coherent scattered light fields can be reconstructed volumetrically. By refocusing the fields to the sample planes, absorption and phase-shift profiles of sparsely distributed samples can be simultaneously inferred in 3D. This holographic advantage is highly useful for spectroscopic imaging of cold atomic samples. However, unlike {\it e.g.} biological samples or solid particles, the quasi-thermal atomic gases under laser-cooling are typically featureless without sharp boundaries, invalidating a class of standard numerical refocusing methods.  Here, we extend the refocusing protocol based on the Gouy phase anomaly for small phase objects to free atomic samples. With a prior knowledge on a coherent spectral phase angle relation for cold atoms that is robust against probe condition variations, an ``out-of-phase'' response of the atomic sample can be reliably identified, which flips the sign during the numeric back-propagation across the sample plane to serve as the refocus criterion.  Experimentally, we determine the sample plane of a laser-cooled $^{39}$K gas released from a microscopic dipole trap, with a $\delta z\approx 1~{\rm \mu m}$$\ll 2\lambda_p/{\rm NA}^2$ axial resolution, with a NA=0.3 holographic microscope at $\lambda_p=770~$nm probe wavelength.
\end{abstract}

\section{Introduction}
In a generic absorption imaging setup, the optical forward scattering $E_s$ from the sample under study is imaged together with the co-propagating probe light $E_p$ onto the imaging sensor arrays. The attenuation of the total intensity $I=|E_p+E_s|^2$ records the in-phase component of $E_s$ relative to $E_p$. Information on the out-of-phase $E_s$ component is lost. Similarly, in phase contrast imaging setups~\cite{Zernike1935, Maurer2008} where $E_p$ is phase-shifted by $\pi/2$, the information loss occurs to the in-phase $E_s$ quadrature. Digital holography~\cite{gabor1948new} recovers the full $E_s$ information by reconstructing the phase of $E_s$ relative to $E_p$ using holograms. For the case of inline holography~\cite{Greenbaum2012, Latychevskaia2019}, the holograms are simply out-of-focus interference fringes between $E_s$ and $E_p$. With the full wavefront knowledge at hand, both the $E_s$ and $E_p$ fields can be volumetrically reconstructed around the sample planes via digital back-propagation. Furthermore, with sufficient  knowledge of the samples, the reconstruction support self-consistent characterization of sparse samples for precise 3D microscopy~\cite{Lee2007, Memmolo2015, Alexander2020}. During the process, to refocus each reconstructed sample image to its respective plane~\cite{Brady2009,Gao2012,Osten2013,ilhan2013autofocusing,wilson20123d,zhang2017edge,Fan2017, Wu2018a, pinkard2019deep} is crucially important. For the purpose, various refocus schemes are developed based on priori knowledge of the samples and their interaction with light. Examples include the methods based on the edge sharpness and sparsity~\cite{ilhan2013autofocusing, Brady2009,zhang2017edge,Fan2017}, the Gouy phase shift~\cite{wilson20123d}, by requiring imaging consistencies under multiple wavelength~\cite{Gao2012} and structured illumination~\cite{Osten2013}, or even by deep learning of complex features~\cite{Wu2018a, pinkard2019deep}. 

\begin{figure*}[htbp]
    \centering
    \includegraphics[width=0.95\linewidth]{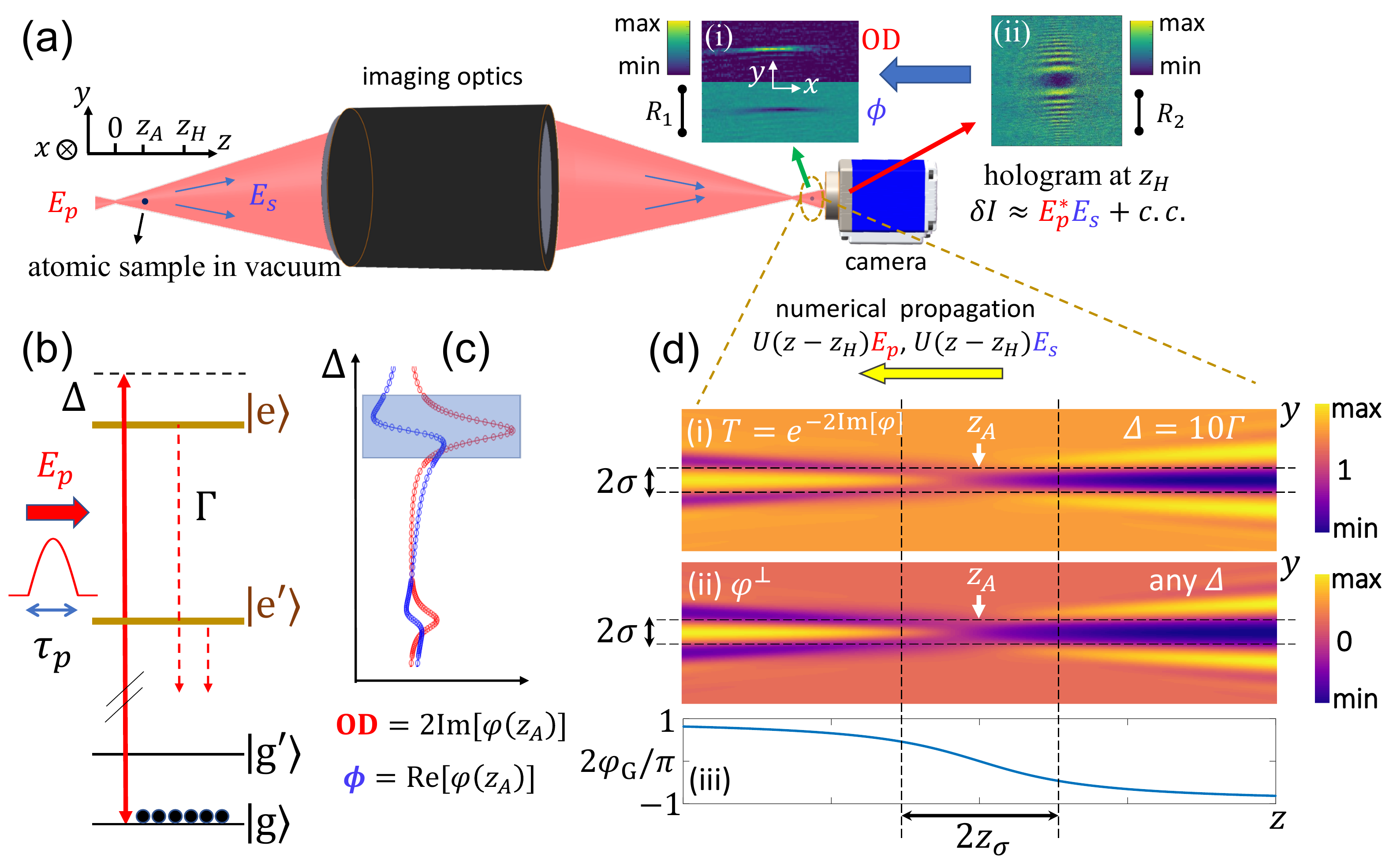}
    \caption{Schematic of the inline holography and the spectroscopic refocus criterion. (a): Experimental setup.  The imaging optics is with magnification $M=1$ and a numerical aperture ${\rm NA}=0.3$. The (a,i) subplots are typical reconstructed optical depth ${\rm OD}$ and phase shift $\phi$ images at the $z=z_A$ plane (experimental data, with peak OD$\approx 0.2$ and $\phi\approx -0.15$). The (a,ii) subplot gives the corresponding reduced hologram $\delta I$ recorded by the camera at $z=z_H$ (peak-to-peak $\delta I\approx 100$ in terms of counts, with photon shot noise at the 20-level~\cite{Wang2022b}.). The scale bars are with $R_{1}=30~{\rm \mu m}$ and $R_{2}=1~{\rm mm}$ respectively.
    (b): A 4-level diagram to represent the light-atom interaction at the $^{39}$K D1 line ($\lambda_p=770~$nm) nearly resonant to the $|g\rangle$ (4S$_{1/2},~F=1$) and $|e\rangle$ (4P$_{1/2},~F'=2$) transition. (c): Simulated {\rm OD} and $\phi$ profiles associated with the D1 interaction under realistic experimental condition according to optical Bloch equations~\cite{Sievers2015}. As detailed in Sec.~\ref{sec:alpha}, the spectroscopic phase angle $\beta_0$ within the shadowed $|\Delta|\leq \Gamma$ follows Eq.~(\ref{eq:tanbeta}), even for strong pulses that saturate the transition. (d,i): Transmission $T(x=0,y,z)\equiv |E_s+E_p|^2/|E_p|^2$ (with the probe detuning $\Delta=10\Gamma$). (d,ii): Out-of-phase response $\varphi^{\bot}(x=0, y,z)$ (at arbitrary $\Delta$) for the simulated D1 sample with $\sigma_y=1.4\lambda_p$ Gaussian profile. Similar to $1-T$, $\varphi^{\perp}$ (Eq.~(\ref{eq:phiPerp})) vanishes at $z=z_A$ and evolves according to the Gouy phase (d,iii) near the $E_s$ center. With a prior $\beta_0$ knowledge by Eq.~(\ref{eq:tanbeta}), suppression of $\varphi^{\perp}(z)$ within $|y|\leq \sigma_y$ serves as refocus criterion to locate $z=z_A$ with the near-resonant probe.  
    }\label{Fig1} 
\end{figure*}

The holographic advantages associated with 3D complex imaging of sparse samples can be highly useful for applications across fields~\cite{Cuche1999,2004microelectromechanical, Bjrn2005Investigation,2008Phase,Greenbaum2012,DeHaan2020}. 
For atomic physics research, over the years efforts have been made for holographic imaging of cold atoms~\cite{Kadlecek2001,turner2005diffraction,Sobol2014,smits2020imaging,Altuntas2021}. In a recent work, we show that an improved holographic technique with suppressed aberration and speckle noises
supports simultaneous retrieval of atomic absorption and phase shift profiles with diffraction-limited spatial resolution and photon shot-noise limited sensitivity~\cite{Wang2022b}. As illustrated in Fig.~\ref{Fig1}a here, the technique uses a precisely pre-characterized probe wavefront $E_p$ to recover the coherent atomic forward scattering $E_s$ with the hologram data (Fig.~\ref{Fig1}a(ii)). Then, both $E_s$ and $E_p$ are numerically propagated from the camera plane $z=z_{H}$ back to the sample plane $z=z_A$ where the 2D optical depth ${\rm OD}(x,y)=-2{\rm Re[log}(1+E_s/E_p)]$ and phase shift ${\phi}(x,y)={\rm Im[log}(1+E_s/E_p)]$  (Fig.~\ref{Fig1}a(i)) are evaluated. Here, similar to the applications in other fields~\cite{Brady2009,Gao2012,Osten2013,ilhan2013autofocusing,wilson20123d,zhang2017edge,Fan2017, Wu2018a, pinkard2019deep}, to localize the atomic plane $z_A$ is crucially important for faithfully retrieving the generic atomic absorption and phase shift properties. However, unlike typical biological or solid samples with sharp boundaries, cold atoms in optical traps typically follow quasi-thermal distributions~\cite{MetcalfBook}, without much distinct features as a priori criterion to perfect the sample-plane refocus. Nevertheless, in early efforts for holographic imaging of cold atoms~\cite{Kadlecek2001,turner2005diffraction,Sobol2014}, the sample-plane localization is still largely based on optimizing certain characteristic spatial features of atomic distribution, with moderate accuracies.

In fact, while cold atomic samples usually lack distinctive spatial structures, there are unique features constrained by atomic physics available for calibrating coherent imaging. For example, in aberration-free in-focus imaging, the power spectrum density of atomic density correlations is expected to be flat for non-correlated atoms~\cite{Hung2011}. The criterion is applied in ref.~\cite{Altuntas2021} to achieve precise refocus of aberrated phase-contrast atomic images. Other than exploiting spatial structures or correlations, in ref.~\cite{smits2020imaging} the authors suggest that for far off-resonant imaging ({\it i.e.}, in Fig.~\ref{Fig1}b the probe detuning $\Delta$ is much larger than the atomic transition linewidth $\Gamma$), that atomic samples appear as phase objects with $|\phi(x,y)|\gg |{\rm OD}(x,y)|$ across the sample becomes a refocus criterion to precisely locate the sample plane. This idea shares the same underlying physics with the refocusing method based on Gouy phase anomaly for 3D localization of small transparent particles~\cite{wilson20123d}. As to be detailed shortly,  relative to the uniform probe wavefront $E_p$, the propagation of spatially confined $E_s$ picks up an extra Gouy phase $\phi_{\rm G}$ in its center in the far field. Therefore, the known relative phase relation between $E_s$ and $E_p$ for phase objects holds sensitively in the near field, naturally serving as a refocus criterion for locating the sample plane. Obviously, this diffraction phase criterion can be generalized for localizing atomic samples probed at arbitrary detuning $\Delta$, if the phase angle  $\beta_0={\rm arg}(\phi+i {\rm OD}/2)$ can be measured and compared with known values precisely. 

In this work, we show that a recently demonstrated phase-angle spectroscopy~\cite{Wang2022b} leads to a robust criterion for locating the sample plane in holographic microscopy of cold atoms, with an achievable axial resolution $\delta z$ well below the diffraction limit. The prior spectral phase angle knowledge exploited in this method, the Eq.~(\ref{eq:tanbeta}) relation to be discussed shortly, is easily understood in the linear optics regime where both ${\rm OD}$ and $\phi$ for a dilute gas can be evaluated analytically~\cite{turner2005diffraction,Meppelink2010}. Practically, to achieve sufficient signals with short and nearly resonant exposures, saturation of atomic transitions can hardly be completely avoided. It is also known that the linear optical response of cold gases is prone to  resonant dipole interactions~\cite{Morice1995,chomaz2012absorption,Zhu2016}. Interestingly, we find that far beyond the linear response regime, the Eq.~(\ref{eq:tanbeta}) relation can hold precisely for isolated atomic transitions probed with smooth and long pulses. Our method thus supports robust localization of atomic sample plane with flexible probe condition managements, for working with denser samples~\cite{chomaz2012absorption} and to achieve photon-shot-noise limited performances~\cite{Sobol2014,Wang2022b}. Experimentally (Fig.~\ref{Fig1}), we demonstrate the spectral refocus method by repeatedly probing an open hyperfine transition of $^{39}$K D1 line with $\tau_p=1~\mu$s pulses, to locate the atomic sample plane with sub-micron axial resolution.  

Previously, the best refocus criterion for imaging cold atomic samples appears to be that based on the atomic shot-noise correlations~\cite{Altuntas2021}. In comparison, our method provides similar accuracy with a much stronger signal for rapid applications. In addition, instead of relying on regularizing contrast transfer functions~\cite{Paganin2002,turner2005diffraction,Altuntas2021}, our holographic method directly supports a large depth of view, with diffraction-limited resolution~\cite{Sobol2014}, for future 3D spectroscopic imaging of sparse atomic samples.




\section{Phase-angle spectroscopy for atomic sample plane localization}\label{sec:Atom}

\subsection{Measurement principles}
As schematically illustrated in the Fig.~\ref{Fig1} setup, we consider holographic imaging of an atomic sample subjected to a spherical probe light $E_p$ illumination at wavelength $\lambda_p$. The atomic sample is centered at ${\bf r}_A=(0,0,z_A)$ with spatial width $\{l_x,l_y\}\ll z_A$, so that the probe light propagates along $z$ through the sample with negligible wavefront curvature itself. We assume thin atomic samples. The length $l_z$ along the light propagation direction satisfies
\begin{equation}
    l_z\ll z_{\sigma}\label{eq:RN}
\end{equation}  
when compared to the Rayleigh distance 
\begin{equation}
    z_{\sigma}=\pi\sigma^2/\lambda_p, \label{eq:zS}
\end{equation}
associated with the smallest spatial feature of interest of the sample characterized by an effective Gaussian width $\sigma$. The spherical $E_p$ in this work is derived from a defocused Gaussian beam. The spherical wave illumination~\cite{gabor1948new} enhances the pixel-resolution and dynamic range of the camera sensors during holographic imaging~\cite{turner2005diffraction, Sobol2014}. Our method can be straightforwardly generalized to plane-wave illumination, as well as structured, complex illuminations.

The total field $E_{\rm tot}=E_s+E_p$ after interacting with the thin sample can be expressed as $E_{\rm tot}=E_p e^{i\varphi}$. The complex phase shift~\cite{Wang2022b}
\begin{equation}
    \varphi(x,y,z)=-i {\rm log}(1+\frac{E_s(x,y,z)}{E_p(x,y,z)})\label{eq:varphi}
    \end{equation}
is related to the optical depth and phase shift as $\varphi(z_A)=\phi+i{\rm OD}/2$.  Notice here and in the following we may omit the $(x,y)$ variable in the 3D distribution functions if no ambiguity is induced.  For a dilute sample of free atoms, the complex numbers are rotated from the real axis by the phase angle $\beta_0={\rm arg}(\varphi(z_A))$ according to the single-atom response. We re-write the complex phase during the backward propagation as 
\begin{equation}
\varphi(x,y,z) =(\varphi^{//}(x,y,z)+i\varphi^{\perp}(x,y,z))e^{i\beta_0}.\label{eq:rot}
\end{equation}
so that, with $\beta(z)={\rm arg}(\varphi(z))$, 
\begin{equation}
    \begin{aligned}
        \varphi^{//}(x,y,z)&= |\varphi(x,y,z)|{\rm cos}(\beta(z)-\beta_0),\\
    \varphi^{\perp}(x,y,z)&= |\varphi(x,y,z)|{\rm sin}(\beta(z)-\beta_0).
    \end{aligned}\label{eq:phiPerp}
\end{equation}
Clearly, the out-of-phase component  $\varphi^{\perp}$ vanishes at $z=z_A$, just like that the attenuation by thin phase objects is zeroed in-focus~\cite{wilson20123d,smits2020imaging}. The effect is illustrated in Fig.~\ref{Fig1}d(i-ii) for the D1 line of $^{39}$K in this work as to be detailed shortly, but is general for arbitrary atomic transitions as long as $\beta_0$ is known for evaluating $\varphi^{\perp}$ with Eq.~(\ref{eq:phiPerp}).

To simplify the following discussion on the propagation effect, we now model the atomic sample by the Gaussian profile with $\sigma_y=\sigma\ll \sigma_x$. Propagating away from $z=z_A$, the defocused $E_s$ near the imaging center (with $|y|\leq\sigma$ in Fig.~\ref{Fig1}c) picks up an additional phase, the Gouy phase $\phi_{\rm G}=-{\rm arctan}\frac{z-z_A}{z_{\sigma}}$ (Fig.~\ref{Fig1}d(iii)), relative to $E_p$~\cite{wilson20123d}. within a region of interest (ROI) defined by $|y|\leq\sigma$, we expect
\begin{equation}
\beta(z)\approx \beta_0-{\rm arctan}\frac{z-z_A}{z_{\sigma}}.\label{eq:beta}
\end{equation}
As to be detailed in the following, the Eqs.~(\ref{eq:varphi}-\ref{eq:beta}) relation can be exploited to locate the $z=z_A$ plane with a given $E_s,E_p$ data set, by minimizing the $\varphi^{\perp}$ components according to a prior $\beta_0$-knowledge.

\subsection{Holographic $E_{s}$, $E_{p}$ reconstruction}\label{sec:rec}
To experimentally evaluate $\varphi(x,y,z)$ according to Eq.~(\ref{eq:varphi}), we need to volumetrically reconstruct $E_{s,p}$ fields from experimental data first. The procedures to infer the probe wavefront $E_p$ from the pre-experimental characterizations, and $E_s$ from the single-shot atomic sample holograms, are detailed in our previous work~\cite{Wang2022b}. Briefly, the probe wavefront at the camera sensor plane $E_p(z_H)=\sqrt{I_p(z_H)}e^{i\phi_p(z_H)}$ is obtained first by a multi-plane Gerchberg-Saxton algorithm~\cite{Gerchberg1972} with multiple $\{I_p(z)\}$ probe intensity measurements as inputs, using a $z-$translating camera. The camera position is then fixed at $z=z_H$ for experimentally recording the holograms with and without the cold atomic samples under study. With careful numerical adjustments and subtractions, the digital images are reduced to represent $I=|E_s+E_p|^2$ and $I_0=|E_p|^2$ respectively. An iterative twin-image removal algorithm~\cite{Sobol2014} is then applied to retrieve $E_s$ from the reduced hologram $\delta I=E^*_p E_s+E_p E_s^*+|E_s|^2$ (Fig.~\ref{Fig1}a(ii)).

With the full wavefront knowledge for $E_s$ and $E_p$ at hand, we numerically propagate both fields from the camera plane $z_H$ to locations $z$ around the sample plane, via the angular spectrum method~\cite{Wang2022b,Zhao2022},
\begin{equation}
    \begin{array}{l}
        E_{s,p}(z)=U(z-z_H)E_{s,p}(z_H),~{\rm with}\\
        U(L)=\hat F^{-1} e^{i\sqrt{k_p^2-k_x^2-k_y^2}L} \hat F.
    \end{array}\label{eq:ASM}
\end{equation}
Here $\hat F$, $\hat F^{-1}$ represents the 2D Fourier transform and the inverse transform respectively: $E(k_x,k_y,z)=\hat F E(x,y,z)$ and $E(x,y,z)=\hat F^{-1} E(k_x,k_y,z)$. $k_p=2\pi/\lambda_p$ is the wavenumber of the probe light.

\subsection{A robust spectroscopic phase angle relation for cold atoms\label{sec:alpha}}
Clearly, to construct a $z=z_A$ refocus criterion with Eqs.~(\ref{eq:rot}-\ref{eq:beta}), prior knowledge of the phase angle $\beta_0$ is essential. 
With the thin sample condition by Eq.~(\ref{eq:RN}), the complex phase shift at $z=z_A$ for a dilute sample is expected to follow the Beer-Lambert law as~\cite{Wang2022b} 
\begin{equation}
\varphi(x,y,z_A)=\frac{1}{2} \int {\rm d}z \varrho(x,y,z) \alpha.\label{eq:varphi0}
\end{equation}
Here $\varrho$ is atomic density distribution and $\alpha$ is the complex atomic polarizability. We therefore expect $\beta_0={\rm arg}(\alpha)$, which is precisely known in the linear optics regime~\cite{Meppelink2010}. However, as suggested in the Introduction, the validity of the linear analysis is practically prone to saturation and resonant dipole interaction effects. 

Beyond the linear analysis, here we generally consider a sample of  multi-level atoms interacting with a probe pulse (Fig.~\ref{Fig1}b). We assume the probe is strong enough so that inter-atomic resonant dipole interaction can be ignored~\cite{chomaz2012absorption}. In light of the fact that the holographic data recorded by the camera is given by $\delta I=\int_0^{\tau_p}\delta I(t)dt$, the effective polarizability $\alpha$ for Eq.~(\ref{eq:varphi0}) is evaluated as
\begin{equation}
    \alpha \approx \frac{\int_0^{\tau_p} {\rm d} t {\bf E}_p^*\cdot \langle{\bf d}\rangle}{\int_0^{\tau_p} {\rm d} t |{\bf E}_p|^2}.\label{eq:alpha}
\end{equation}
Furthermore, when the probe frequency $\omega_p$ is nearly resonant to an isolated $|g\rangle-|e\rangle$ transition, the induced complex dipole moment can be approximated as 
\begin{equation}
    \langle {\bf d}(t) \rangle\approx \rho_{eg}(t){\bf d}_{ge}e^{-i\omega_p t}.   \label{eq:d}
\end{equation}
Here ${\bf d}_{eg}$ is the dipole matrix element of the $|g\rangle-|e\rangle$ transition. The coherence $\rho_{eg}(t)$ obeys the multi-level master equation for the atomic density matrix $\rho$~\cite{ScullyBook}, as detailed in Appendix~\ref{sec:secular}. 

With the key Eqs.~(\ref{eq:alpha})(\ref{eq:d}) assumptions, in Appendix~\ref{sec:secular} we show that for a smooth pulse with duration $\tau_p\gg 1/\Gamma$ at detuning $|\Delta|\leq\Gamma$ , $\beta_0$ is approximated by
\begin{equation}
    \beta_0 (\Delta)={\rm arccot}\left(2a \frac{\Delta}{\Gamma}+b\right) \label{eq:tanbeta}
\end{equation}
with very good accuracy. For example, for the D1 line probed in this work as to be detailed in Sec.~\ref{sec:exp} (Fig.~\ref{Fig1}c), deviation from Eq.~(\ref{eq:tanbeta}) within $|\Delta|\leq \Gamma$ is less than $1\%$ in terms of minimal $\varphi^{\perp}/|\varphi|$ by Eq.~(\ref{eq:phiPerp}). Here $a=-\frac{\Gamma}{\Gamma+\gamma}$ and $b=-\frac{\delta}{\Gamma+\gamma}$ parametrize the average Stark shift $\delta$ and optical pumping rate $\gamma$ induced by all the $|g\rangle -|e'\rangle$ and $|g'\rangle -|e\rangle$ couplings that off-resonantly mix the atomic states during the $\tau_p$ time. The spectral isolation here requires the transition frequency $\omega_{e g}$ to be far away from these ground- or excited-state sharing transitions, as well as all the other $|g'\rangle-|e'\rangle$ transitions: $|\omega_{e'g'}-\omega_{eg}|,|\omega_{e e'}|\gg \Gamma+\Gamma'$, $|\omega_{g g'}|\gg \Gamma$. The probe pulse can be fairly short and strong, as long as these other transitions are only excited perturbatively.  The robustness of the phase-angle relation by Eq.~(\ref{eq:tanbeta}) makes it particularly convenient to constrain the sample plane $z_A$ of dilute atomic gases in digital holography. 




\subsection{Spectroscopic sample plane localization}\label{sec:loc}

Experimentally, as to be detailed in Sec.~\ref{sec:exp}, a set of holograms for a standard, free-space sample of dilute atoms is recorded in repeated preparation-measurement cycles at various detuning $\{\Delta_j\}$ with $|\Delta_j|\leq \Gamma$. The complex phase shifts $\{\varphi_j\}$ are then evaluated according to Eq.~(\ref{eq:varphi}). To locate the atomic sample plane, we  numerically propagate $E_{s,p}(z)$ according to Eq.~(\ref{eq:ASM}) to minimize the cost function
\begin{equation}
    \begin{aligned}
L(z;\{\varphi_j,\Delta_j\})&=\sum_j (\overline{\varphi_j^{\perp}}(z))^2\\
&=\sum_j \left ({\rm sin}[\beta_0(\Delta_j)]{\rm Re}[\overline{\varphi}_{j}(z)]-{\rm cos}[\beta_0(\Delta_j)]{\rm Im}[\overline{\varphi}_j(z)]\right )^2
    \end{aligned}\label{eq:cost}
\end{equation}
which implicitly depends on the lineshape parameters  $a,b$ through Eq.~(\ref{eq:tanbeta}), as well as the sample plane $z_A$ through Eq.~(\ref{eq:beta}). Each out-of-phase component $\overline{\varphi^{\perp}_j}(z)$ by Eq.~(\ref{eq:phiPerp}) is averaged within the ROI defined by $|y|<\sigma$ (Fig.~\ref{Fig1}d).   With the $L$-minimization, we expect a linear relation between ${\rm OD}$ and ${\phi}$ within $|\Delta|\leq \Gamma$:
\begin{equation}
\left(\frac{{\rm Re}[\overline{\varphi}(z)]}{{\rm Im}[\overline{\varphi}(z)]}\right)_{z=z_A}=2 a\frac{\Delta}{\Gamma}+b.
 \label{eq:odPhiL}
\end{equation}

To understand the accuracy of the $L-$minimization by Eq.~(\ref{eq:cost}), we notice the $\overline{\varphi^{\perp}}_j$ data entering the analysis is fundamentally limited by the uncertainty of the phase angle $\beta(z)={\rm arg}(\varphi(z))$ retrieved from the hologram data. In particular, with $\beta_j\approx \beta_0(\Delta_j)$, the noise in  the inferred $\overline{\varphi^{\perp}_j}$ at each detuning $\Delta_j$ is photon-shot-noise-limited to $\delta\overline{\varphi^{\perp}_j}=|\overline{\varphi_j}|\times  \delta\beta_j$ as~\cite{Wang2022b}
\begin{equation}
\delta\beta_j = \frac{1}{\sqrt{N_{s,j}}},\label{eq:db}
\end{equation}
with $N_{s,j}\propto \sum_{\rm ROI} |E_s^{(j)}|^2$ summing the elastically scattered photons from the ROI to the camera. 

The photon shot noise affects the predictions to all the $a,b,z_A$ parameters. However, with $\Delta_j$ uniformly sampling $|\Delta|\leq \Gamma$ so that $a,b$ can typically be fixed quite precisely, we may simplify the analysis by ignoring the correlations so that $\delta L=0$ suggests $\sum_j\delta\overline{\varphi^{\perp}}_j+\delta z_A \partial_{z_A} \overline{\varphi^{\perp}_j}=0$. Together with Eq.~(\ref{eq:beta}), we arrive at photon-shot-noise limited axial resolution
\begin{equation}
    \delta z_A= \eta \frac{z_{\sigma}}{\sqrt{N_s}}.\label{eq:res}
\end{equation}
Here $N_s=\sum_j N_{s,j}$ is the number of all the elastically scattered photons entering the data analysis. The $\eta$ is a sample shape dependent factor, with $\eta=1$ for the Gaussian shaped sample. 

We now discuss the choice of ROI for the coherent signal averaging in Eqs.~(\ref{eq:cost})(\ref{eq:odPhiL}). As shown numerically in Fig.~\ref{Fig1}d(ii) (Appendix~\ref{sec:simu}), the rapid sign inversion for $\varphi^{\perp}$ along the light propagation direction $z$  is most pronounced near the $E_s$ center with $|y|<\sigma$ in the plot. With a large $\Delta z=z-z_A$ to be comparable to $z_{\sigma}$ (Eq.~(\ref{eq:zS})), oscillatory $\varphi^{\perp}(y,z)$ is developed along $y$ due to the curvature mismatch between $E_s$ and $E_p$. Therefore, for the $\sigma$-sized atomic sample, $|y|\leq\sigma$ is a natural ROI choice to isolate the uniform $\varphi^{\perp}(z)$ center for the $\overline{\varphi^{\perp}}(z)$ average. A too large ROI would reduce the $z$-dependence of $\overline{\varphi^{\perp}}(z)$ comparing to Eqs.~(\ref{eq:phiPerp})(\ref{eq:beta}). Conversely, a too small ROI would reduce the total number of elastically scattered photons $N_s$ entering the data analysis, compromising the photon shot noise limit (Eq.~(\ref{eq:res})). Practically, the atomic samples cannot always be approximated by Gaussian profiles. The size $\sigma$ may not always be assumed as prior knowledge either. 
For the small samples to be discussed in this work (Fig.~\ref{Fig1}a), the ROI should in principle be refined toward $|y-y_0|<\sigma$  with a suitable central position $y_0$ and width $\sigma$ to balance the $z-$sensitivity of the $L$-function in Eq.~(\ref{eq:cost}) with the amount of ROI-photons $N_s$. Practically, with the sample plane emphatically determined, we find that simply by thresholding the approximately refocused $|E_s|^2$ intensity, {\it e.g.} ROI=1 for $|E_s|^2>\epsilon |E_s|^2_{\rm max}$, $\epsilon\sim 0.1$, then nearly optimal $\delta z$ sensitivity for the Eq.~(\ref{eq:cost}) minimization can be achieved.

\subsection{Correcting high-order aberrations}\label{sec:aber}

From Eqs.~(\ref{eq:cost})(\ref{eq:res}), the high quality atomic sample plane localization relies on high quality minimization of the $L-$function toward the photon shot noise limit. The fit quality is affected by the imperfect optical system itself. For lensless holographic imaging~\cite{Sobol2014}, the optical transfer function according to free-space propagation is easily modeled. However, in most cold atom experiments, the imaging system usually requires an optical train to relay the coherent wavefronts from the samples in vacuum to the camera outside the vacuum~\cite{Alexander2020, Wang2022b,Altuntas2021}, as schematically illustrated in Fig.~\ref{Fig1}a. Even for perfect optics, high-order aberration  correction is required for volumetric imaging across a large depth of view.

Here, the high-order aberration correction can be achieved simultaneously with the $z=z_A$ sample plane localization, through the minimization of $L-$function by Eq.~(\ref{eq:beta}). For example, this can be achieved by setting up numerical Zernike plates~\cite{Altuntas2021} at $z=z_H$ with coefficients entering the $L-$function for the optimization. This is a step left for future work.


\begin{figure}[htbp]
    \centering
    \includegraphics[width=.8\linewidth]{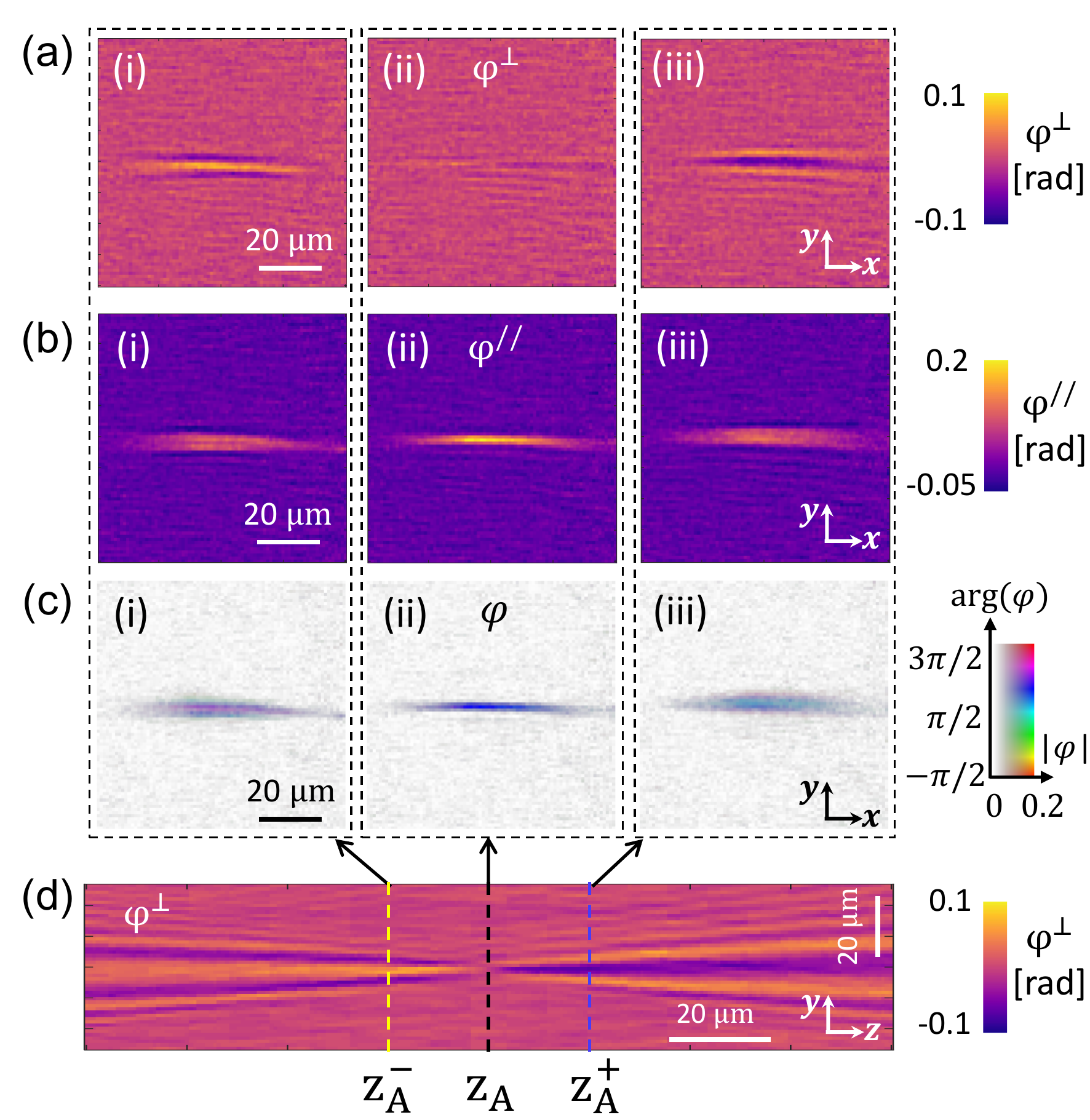}
    \caption{Holographically reconstructed images of a free $^{39}$K gas during the sample-plane refocus. The probe detuning is $\Delta=\Gamma$ in this example. The out-of-phase $\varphi^{\bot}(z)$ and the in-phase $\varphi^{//}(z)$ are displayed in (a)(b) respectively. A color domain plot of $\varphi$ is given in (c), with the $|\varphi|$ strength and the phase angle $\beta={\rm arg}(\varphi)$ encoded by brightness and color respectively. The images in (i)(ii)(iii) are evaluated at $z=z_A^-,z_A,z_A^+$ planes respectively, with $z_A^{\pm}=z_A \pm 20~{\rm \mu m}$.  The $\varphi^{\perp}$ plot in (d) is similar to Fig.~\ref{Fig1}(d,ii), but is reconstructed from the experimental data.}
    \label{Fig:single} 
\end{figure}

\section{Experimental demonstration}\label{sec:exp}
\subsection{Methods}
Our experiment demonstration is based on a $^{39}$K holographic microscope on the D1 line. To facilitate the following discussions, in Fig.~\ref{Fig1}b we refer the $4S_{1/2}$, $F=1,2$ hyperfine ground states as $|g\rangle,|g'\rangle$, and $4P_{1/2}$, $F'=1,2$ excited states as $|e'\rangle,|e\rangle$ respectively~\cite{Weller2015}. As schematically illustrated in Fig.~\ref{Fig1}a, an ${\rm NA}=0.3$ optical train with magnification $M=1$ relays the probe wavefront $E_p$ and the forward scattering $E_s$ by the cold atomic sample to the digital camera. 
The probe wavelength  {$\lambda_p=770~$nm} is nearly resonant to the $|g\rangle-|e\rangle$ hyperfine transition with a natural linewidth $\Gamma=2\pi\times 5.96$~MHz. The transition is spectrally isolated from $|g\rangle-|e'\rangle$ and $|g'\rangle-|e\rangle$ transitions by $\omega_{e,e'}=2\pi\times 55.5~$MHz and $\omega_{g,g'}=2\pi\times461.7~$MHz respectively. Up to $10^3$ atoms are laser-cooled to a temperature of tens of micro-kelvin~\cite{Sievers2015} and loaded into a microscopic optical dipole trap (ODT) composed by a focused $\lambda'=780~$nm laser along $x$. The approximately Gaussian-shaped atomic sample is with $\sigma_x = 15~{\rm \mu m}$ along $x$. The width $l_{y,z}$ estimated to be less than 1~${\rm \mu m}$ is below the $(\delta y)_{\rm res}=\lambda_p/{\rm NA}= {2.6\ \mu m}$ diffraction limit~\cite{Sobol2014a} of the NA=0.3 holographic microscope~\cite{Zhao2020}. In absence of imaging aberrations, we expect the diffraction-limited sample images to have an apparent Gaussian width $\sigma_y\approx (\delta y)_{\rm res}/2\sqrt{2}$ along $y$ of about $\sigma_y= {1.0~{\rm \mu m}}$. The associated diffraction distance $4 z_{\sigma}\approx 16~{\rm \mu m}$ (Eq.~(\ref{eq:zS})) is close to the diffraction-limited imaging axial resolution $(\delta z)_{\rm depth}=2\lambda_p/{\rm NA}^2$. 

To spectroscopically locate the atomic sample plane,  a sequence of two images $I_{1,2}^{(j)}$ are recorded at each probe detuning $\Delta_j$ with and without the atomic sample in repeated measurements. The CCD camera with $1040\times 1392$ pixels, each $6.45\times 6.45~{\rm \mu m}$$^2$ in size, is effectively placed at $z_H =10.4$~mm to record holograms of the atomic sample at $z_A=0.7$~mm.  The camera exposure is set as $\tau_e=1$~ms. As detailed in Appendix~\ref{sec:expDetail}, to avoid inhomogeneous light shifts that tend to invalidate Eq.~(\ref{eq:odPhiL}), these ``standard samples'' are released from ODT before the holographic imaging.  Notice the $|g\rangle - |e\rangle$ transition is open: during the probe excitation, spontaneous $|g\rangle\rightarrow |e\rangle~{\rm or}~|e'\rangle \rightarrow |g'\rangle$ Raman scattering tends to quench the atomic population into the dark $|g'\rangle$. Therefore, instead of probing the atoms continuously, we interleave a train of $\tau_p=1~\mu$s probe pulses with 4~$\mu$s of trapping+cooling pulses composed of the ODT beam with a D1 molasses~\cite{Sievers2015} blue detuned from $|g'\rangle-|e\rangle$ transition, which not only help to maintain the samples' shape, location, and temperature, but also depump the internal states back to $|g\rangle$ for the nearly resonant imaging.  The probe pulse is set at a moderate $I\approx 3$~mW/cm$^2$ intensity in this work so that the near-resonant ${\rm OD}$ and $\phi$ are significant for the microscopic and dilute samples. Since the completion of this work, we have verified that our method works well at higher intensity, both numerically (Appendix~\ref{sec:secular}) and experimentally~\cite{Wang2022b}.

We follow the procedure outlined in Sec.~\ref{sec:rec} to reconstruct $E_{s,p}^{(j)}$ at each probe detuning $\Delta_j$, using the $I_{1,2}^{(j)}$ hologram data taken from repeated measurements. Depending on the desired signal to noise ratio, $N_{\rm ave}=2\sim 15$ holograms obtained at a same measurement condition are averaged to $I_{1,2}^{(j)}$ before proceeding the $E_{s}^{(j)}$ reconstructions. We then propagate both fields according to Eq.~(\ref{eq:ASM}) to  retrieve $\varphi_j(z)$ across the sample plane $z_A$, for the $L$-function minimization (Eq.~(\ref{eq:cost})). The performance of the spectroscopic refocus method is evaluated by repeating the measurements, typically within an hour of measurement time, to check the consistency of the predicted $z=z_A$ values. In addition, the results are compared with the more traditional method based on minimizing the apparent sample width, given by
\begin{equation}
l_y(z)=4\sqrt{\overline{y^2}-\overline{y}^2}.\label{eq:ly}
\end{equation}
Here $\overline {y}, \overline {y^2}$ are the 2D average of $y,y^2$ in the $x-y$ plane, weighted by the 2D complex phase magnitude $|\varphi(x,y,z)|$ within a large enough ROI$'$, during the numerical $z-$scan of $E_{s,p}$. Experimentally we retrieve $l_y(z)$ by fitting $x-$ averaged $|\varphi(x,y,z)|$ with 1D Gaussian profiles. The factor of 4 is chosen so that $l_y=2\sqrt{2}\sigma_y$ for the Gaussian beam model.





\begin{figure}[htbp]
    \centering
    \includegraphics[width=.7\linewidth]{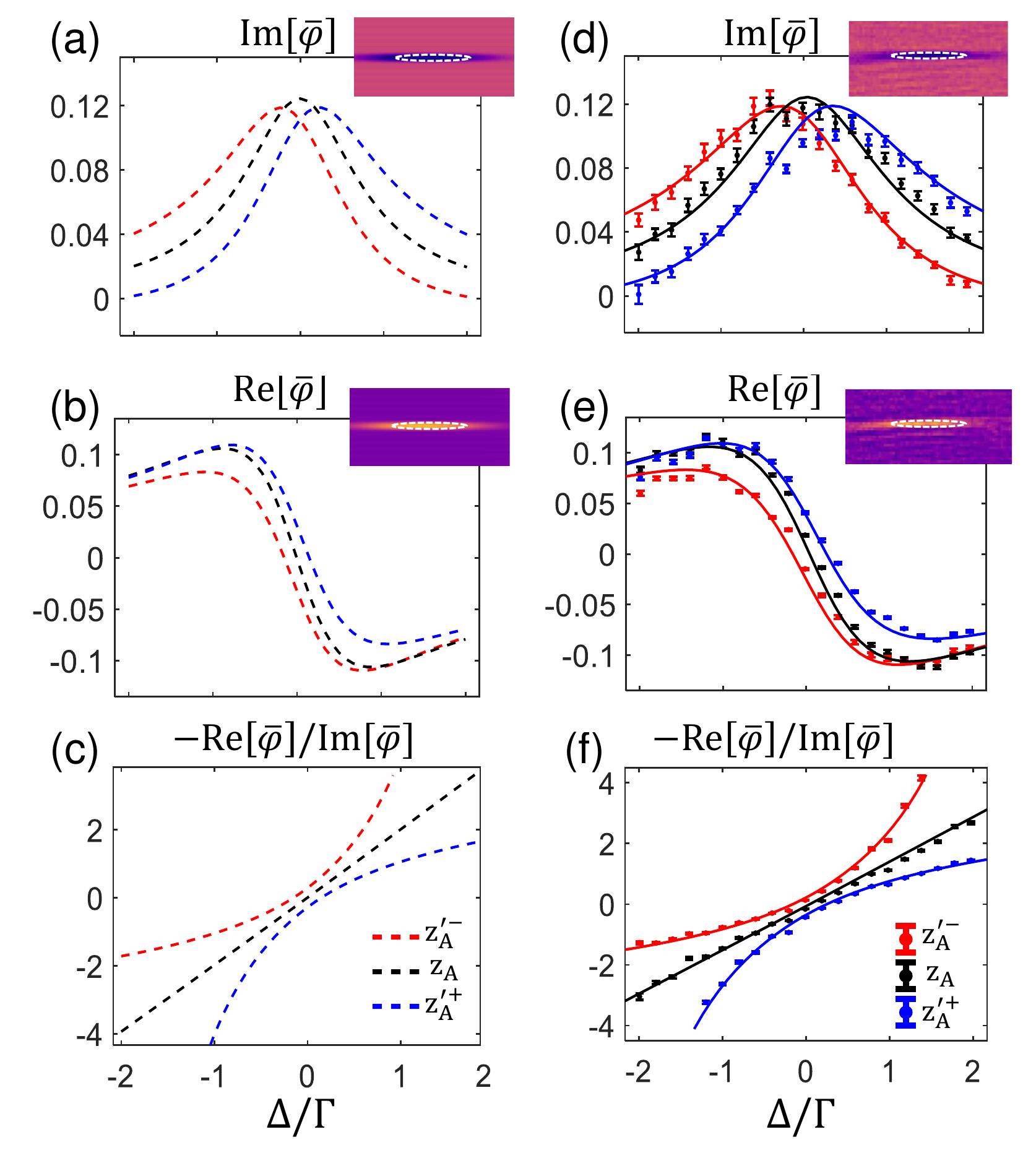}
    \caption{Spectroscopic signature of the ROI-averaged complex phase shift $\overline{\varphi}(z)$ during the refocus to the atomic sample plane. The insets of (a,b,d,e) provide the ${\rm Im}(\varphi(z))$ and ${\rm Re}(\varphi(z))$ images close to $z=z_A$, with ROI marked with dashed circles. (a-c) are according to numerical simulation of 2-level atoms (Appendix~\ref{sec:simu}), with sample parameters adjusted according to the experimental situation detailed in Sec.~\ref{sec:exp}. The experimental data are presented in (d-f) with scattered symbols, where the solid lines are from 2-level atom model with $a=-0.84, b=0.08$. Here $z_A^{'\pm}=z_A\pm10~{\rm \mu m}$. In both (c)(f), the  ${\rm Re}[\varphi(z)]/{\rm Im}[\varphi(z)]$ ratio appears straight only at $z=z_A$ (Eq.~(\ref{eq:odPhiL})). }
    \label{Fig:spec} 
\end{figure}

\subsection{Results}

We first present typical 2D $\varphi(z)$ profiles in Fig.~\ref{Fig:single} around the sample location $z=z_A$. The probe detuning is chosen as $\Delta=\Gamma$ in this example with substantial optical peak depth ${\rm OD}\approx 0.2$ and peak $\phi \approx -0.15$ respectively (see Fig.~\ref{Fig1}a(i)). To improve the display, the holographic data is averaged over $N_{\rm ave}=14$ images (see Fig.~\ref{Fig1}a(ii)). We rotate the complex $\varphi$ with the known $\beta_0(\Delta)$ according to Eq.~(\ref{eq:rot}), which, according to the $L$-minimization to be described shortly, is adjusted to be ${\rm arctan}(2a+b)$ with $a=-0.84, b=0.08$ (fit from experiment data, see Table.~\ref{Table:cost} "average" column). In Fig.~\ref{Fig:single}a we see the $\varphi^{\perp}$ almost vanishes at $z=z_A$ while some weak fringes are still seen, due to the uncompensated high-order aberrations (Sec.~\ref{sec:aber}). On the other hand, substantial $\varphi^{\perp}(z)$ are developed at $\Delta z=\pm 20~{\rm \mu m}$. In Fig.~\ref{Fig:single}d the $\varphi^{\perp}(x=0,y,z)$ similar to Fig.~\ref{Fig1}d(ii) is given, where we see the experimental data matched very well with the theoretical expectation. The complex phase shift $\varphi(z)$ is given in Fig.~\ref{Fig:single}c with the color-domain plots. At the precisely refocused $z=z_A$, the $\varphi(z)$  becomes ``monomorphous'' with a uniform $\beta={\rm arg}(\varphi(z_A))$ distribution~\cite{turner2005diffraction}, as expected.

Next, the highly $z$-sensitive Eqs.~(\ref{eq:cost})(\ref{eq:odPhiL}) criterion is illustrated in Fig.~\ref{Fig:spec} with the ROI-averaged $\overline{\varphi}(z)-\Delta$ curves. Here we still have $N_{\rm ave}=14$. The 2D $\varphi(z)$ distribution around $z=z_A$ as those in Fig.~\ref{Fig:single} are reproduced in the inset plots of Fig.~\ref{Fig:spec}(a,b,d,e). We evaluate  $\overline{\varphi}(z)$ as described in Sec.~\ref{sec:loc}, within the ROI that are marked with dashed circles in the insets. The Fig.~\ref{Fig:spec}(a-c) data are according to the experimental geometry, but for linear response of ideal 2-level atoms ($a=-1,b=0$). The solid curves in Fig.~\ref{Fig:spec}(d-f) are instead numerically generated by adjusting the saturation of the 2-level model (Appendix~\ref{sec:simu}) together with $a=-0.84, b=0.08$ parameters to fit the experimental data. In Fig.~\ref{Fig:spec}(c,f) we see the linearity of the phase angle is strongly impacted by the deviation from the sample plane $z_A$ by a distance as small as $\Delta z\approx \pm  10~\mu$m to be comparable to $2 z_{\sigma}$, in agreement with Eq.~(\ref{eq:beta}).


\begin{figure}[htbp]
    \centering
    \includegraphics[width=1\linewidth]{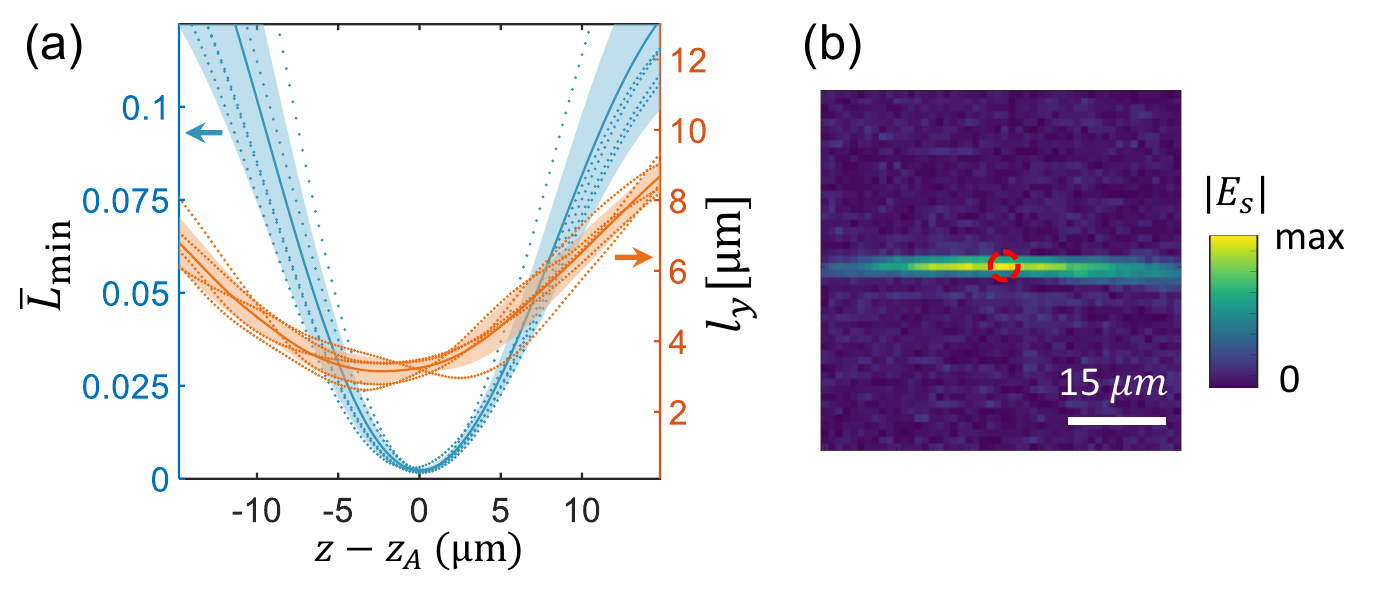}
    \caption{(a): Refocus $^{39}$K sample to $z=z_A$ plane by minimizing the spectroscopic $\overline{L}_{\rm min}(z)$ (Eq.~(\ref{eq:Lr}), blue) and apparent width $l_y(z)$ (Eq.~(\ref{eq:ly}), orange) with $z-$scan near $z_A$.  Five data sets, with $N_{\rm ave}=3,3,3,3,2$ hologram-average respectively, are shown with dot symbols. The solid lines give average $\overline{L}_{\rm min}$ and $l_y$. The shadows provide standard deviations. (b): A typical in-focus $|E_s|$ image with $N_{\rm ave}=14$ hologram-average, where the red circle suggests the diffraction limit $(\delta y)_{\rm res}=\lambda_p/{\rm NA}$. }
     \label{Fig:cost} 
\end{figure}


Having introduced the general spectroscopic features of the reconstructed $\varphi(z)$ near $z=z_A$, we now present details of the $z_A$-plane localization by minimizing the $L-$ function in Eq.~(\ref{eq:cost}). Specifically, for a set of 25 $\{\varphi_j,\Delta_j\}$ data with $\Delta$ scanning from -6~MHz to 6~MHz by 0.5~MHz steps, we globally minimize  $L$ to obtain the $a_{\rm opt},b_{\rm opt},z_{A}$ parameters. We then plot the normalized cost function, 

\begin{equation}
    \overline{L}_{\rm min}(z)\equiv L(z,a_{\rm opt},b_{\rm opt})/\sum_j|\overline{\varphi}_j|^2,\label{eq:Lr}
\end{equation}
in Fig.~\ref{Fig:cost}a vs $z$. To check the consistency of the localization, the same procedure is repeated with five data sets, with holograms in each set averaged by a moderate $N_{\rm ave}=2\sim 3$.  The $\overline{L}_{\rm min}(z)$ curves are compared with the apparent width $l_y(z)$ according to Eq.~(\ref{eq:ly}) evaluated with the same data set. The detailed numbers are given in Table~\ref{Table:cost}. For a clear comparison, in both Fig.~\ref{Fig:cost}a and Table~\ref{Table:cost} the distance $z$ is evaluated relative to the average-$z_A$ from the five $L_{\rm min}-$based predictions. 

Experimentally, from Fig.~\ref{Fig:cost}a we see $\overline{L}_{\rm min}(z)$ on the displayed vertical scale almost vanishes. Indeed, $\overline{L}_{\rm min}(z_A)\approx 0.2\%$ (Table~\ref{Table:cost}) is less than a tenth of $\overline{L}_{\rm min}(z_A\pm \Delta z)$ by a slight defocusing distance $\Delta z=5~{\rm \mu m}$.  For comparison, $l_y(z)$ hardly changes by $30\%$ by the same defocus. As by Line~2 of Table~\ref{Table:cost}, the $z=z_A$ localization in repeated measurements show remarkable consistency with an estimated standard deviation of $\delta z_A=0.3~\mu$m.  For comparison, the sample width method in Fig.~\ref{Fig:cost}a (orange lines) and Line~(5-6) of Table~\ref{Table:cost} show a substantially larger $\delta z_A=2~\mu$m. The much better performance by the spectroscopic method is associated with the aforementioned strong $z-$dependence of the cost function $L_{\rm min}(z)$, due to the Gouy phase anomaly (Eq.~(\ref{eq:beta})), which makes the spectroscopic method substantially more resilient to imaging noises.

Ideally, according to Eq.~(\ref{eq:db}), we expect $\overline{L}_{\rm min}(z_A)=1/N_s$ in the photon-shot-noise limit. for the Fig.~\ref{Fig:cost}a and Table~\ref{Table:cost} results, the total number of elastically scattered photons in each data set of holograms at the 25  $\Delta_j$ detuning is estimated to be $N_s\sim 10^6$, taking into account the $\sim 20\%$ quantum efficiency of our camera (Pico Pixelfly). However, as in Fig.~\ref{Fig:single}a our imaging system is not ideal so that the observed $\overline{L}_{\rm min}(z_A)\approx 2\times 10^{-3} \gg 1/N_s$ is instead limited by high-order aberrations. 
The observed $\overline{L}_{\rm min}(z_A)$ is also substantially larger than those caused by deviation from the Eq.~(\ref{eq:tanbeta}) relation, due to transient and multi-level effects (Appendix~\ref{sec:secular}) which is numerically estimated at a $10^{-5}$ level~\cite{foot:XH}. In other words, there is an intrinsic uncertainty to the phase angle $\delta \beta_0\approx (\overline{L}_{\rm min}(z_A))^{1/2}\approx 0.05$, due to the imaging system smearing itself. With $\overline{L}_{\rm min}(z_A+\Delta z)=\overline{L}_{\rm min}(z_A)+\xi \Delta z^2/2$ in quadratic form, we attribute the observed ${\overline L}_{\rm min}(z_A)$ as unreliable modeling that limits the absolute axial resolution  to $\delta z_{A}=\sqrt{\overline{L}_{\rm min}(z_A)/\xi}\approx 1.0~{\rm \mu m}$. Finally, it is useful to remark that although the data in Table~1 suggests any optical drifts between the atomic sample and camera is small in repeated measurements, practically any drifts during the $\Delta_j$-scan measurements effectively increase the sizes of the ``average samples'' entering the data analysis. In that case, our $L_{\rm min}$ method should still operates to find the average $z_A$ planes.

\begin{table}[h]  
    \centering 
	\scalebox{0.9}
{
    \begin{tabular}{c c c c c c c c} 
    \hline\hline   
      & repeat1 & repeat2 & repeat3 & repeat4 & repeat5 & average & std 
    \\ [0.5ex]  
    \hline 
    $\overline{L}_{\rm min}(z_A)\times 10^{3}$ & 2.79 & 1.55 & 2.41 & 1.98 & 1.79 & 2.10 & 0.44 \\[1ex]  
    
    $z_A$ [$\mu$m] & -0.6 & 0.1 & 0.3 & 0.2 & -0.1 & -0.02 & 0.32 \\[1ex]  
    
    $a_{\rm opt}$ & -0.86 & -0.85 & -0.86 & -0.83 & -0.84 & -0.84 & 0.017  \\[1ex]  
    
    $b_{\rm opt}$ & 0.048 & 0.088 & 0.090 & 0.111 & 0.075 & 0.083 & 0.021 \\[1ex]  
    
    \hline   
    
    $l_{y} (z_A) $ [$\mu$m] & 2.96 & 2.78 & 2.62 & 3.37 & 3.38 & 3.02 & 0.31 \\[1ex]  
    
    $z_A$ [$\mu$m] & 2.5 & -2.6 & -3.3 & -2.1 & -2.2 & -1.54 & 2.06 \\[1ex]

    \hline \hline 
      
    \end{tabular}  
    }
      \caption{Comparison of the spectroscopic $L-$minimization (top) and sample-width $l_y$-minimization (bottom) for the $z=z_A$ localization. std= standard deviation.} 
    
    \label{Table:cost} 
    \end{table}

\section{Discussions}\label{sec:summary}

The last twenty years witness rapid developments of quantitative imaging techniques in digital holographic microscopy, with applications across fields~\cite{Cuche1999,2004microelectromechanical, Bjrn2005Investigation,2008Phase,Greenbaum2012,DeHaan2020}. In comparison, holographic imaging for atomic physics research has been underdeveloped. A list of unique technical challenges needs to be addressed~\cite{Sobol2014a,Wang2022b}, before the holographic method can be applied with sufficient accuracy for imaging the highly fragile ultra-cold atomic samples. This work aims to resolve a particular challenge: the precise localization of the sample plane for retrieving the generic optical response of the atoms. The difficulty arises from the fact that typical atomic samples are spatially featureless. Previously, the only effort to address the problem appears to be exploiting atomic shot-noise correlations in phase-contrast imaging~\cite{Altuntas2021}. 

In this work, instead of relying on spatial information to form refocus criterion, we propose to utilize characteristic spectroscopic features of atomic transitions for precise refocus in holographic microscopy. The underlying principle is to exploit the additional diffraction phase in the forward direction picked up by small objects, known as Gouy phase anomaly~\cite{wilson20123d}, that leads to deviation of apparent spectroscopic responses from those predicted by theory. The idea has already been demonstrated for localizing transparent objects~\cite{wilson20123d,smits2020imaging}. We combine the diffraction phase idea with the unique ability of holographic microscopy for resolving the complex phase shift~\cite{Wang2022b}, and propose a spectroscopic criterion to robustly localize the atomic sample plane. The proposal not only utilizes the fact that for dilute, thin samples the spectral phase angle is insensitive to atomic density fluctuation~\cite{Wang2022b}, but also exploit an interesting phase-angle relation (Eq.~(\ref{eq:tanbeta})) which holds precisely for multi-level atom driven by strong optical pulses (Appendix~\ref{sec:secular}). 

Experimentally, this work demonstrates super-resolved sample plane localization during digital holography of a diffraction-limited, laser-cooled $^{39}$K sample with sub-micron  repeatability. This axial resolution is improved from the traditional method based on fitting the sample widths (Fig.~\ref{Fig:cost}a) by nearly an order of magnitude, in presence of imaging noises in our system. The absolute axial resolution of $\delta z_A\approx 1~\mu$m is yet limited by high-order aberrations of the imaging system itself (Sec.~\ref{sec:exp}), that can be minimized in future work (Sec.~\ref{sec:aber}). With the improvements, we expect the absolute axial resolution to reach the sub-micron level too, to be even smaller than the sample size itself. Our method can be applied to larger samples, where the atomic density fluctuations~\cite{Hung2011,Altuntas2021} lead to the required diffraction phase shifts. By properly choosing a set of ROIs (Eq.~(\ref{eq:cost})) (Sec.~\ref{sec:loc}), spectroscopic signatures of density-fluctuating features at various length scales of interest can be exploited to efficiently locate the central planes of the samples with the holographic microscope. Finally, it is important to note that in our experiment, the peak atomic density of about $10^{13}{\rm cm}^{-3}\ll 1/k_p^3$ is quite dilute, while the peak optical depth ${\rm OD}_{\rm max}<0.5$ is still small. To exploit our method for localizing samples with higher OD and density, stronger pulses should help to suppress contribution of resonant dipole interactions that would otherwise modify the line shape~\cite{chomaz2012absorption, Zhu2016} to compromise the Eq.~(\ref{eq:tanbeta}) criterion.

With the precise knowledge of the sample plane location, the complex spectroscopy method in this work can be uniquely powerful for resolving the phase angle information of ultra-cold samples next~\cite{Wang2022b}. In particular, to infer nontrivial, correlated optical responses of high OD, high density gases~\cite{chomaz2012absorption, Zhu2016}, the atomic sample plane can be spectroscopically located by probing a strong, isolated atomic transition with strong enough pulses first, as in this work. The phase-angle spectroscopy of the cooperative responses of the denser samples can then be reliably retrieved, in presence of density fluctuations generic to cold atomic samples which typically prevent regular imaging methods from obtaining accurate spectroscopic information in single-shots~\cite{Marti2018,
Li2020Wu}. Our spectroscopic method can also be extended to locate multiple samples in digital holography, where the highly precise sample plane localization forms an excellent starting point for complex spectroscopic imaging~\cite{Wang2022b,foot:XH} of sparsely distributed cold atomic samples in 3D~\cite{Nelson2007, Barredo2018}.

\section*{Funding information}
The authors acknowledge support from National Key Research Program of China (2022YFA1404200; 2017YFA0304204); NSFC under Grant No.~12074083; and the Original Research Initiative at Fudan University.

\section*{Disclosures}
The authors declare no conflicts of interest.

\section*{Data availability}
Data underlying the results presented in this paper are not publicly available at this time but may be obtained from the authors upon reasonable request.

\section*{Acknowledgments}
We thank Dr. Liyang Qiu for helping on numerical calculations in this project.

\appendix

\section{Derivation of Eq.~(\ref{eq:tanbeta})}\label{sec:secular}
\begin{figure}[htbp]
    \centering
    \includegraphics[width=.7\linewidth]{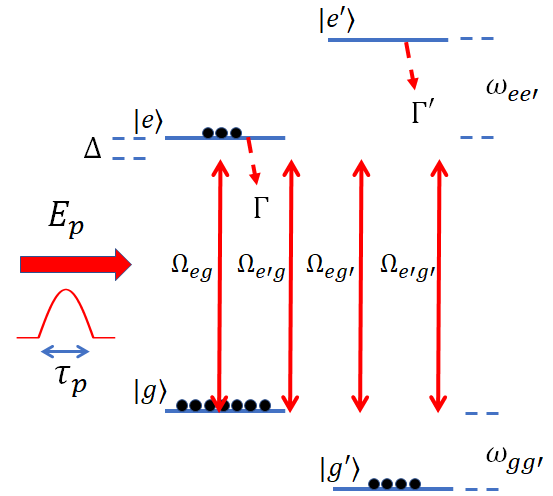}
    \caption{Schematic of a multi-level atom interacting with a monotonic probe $E_p$ with duration $\tau_p$. The probe frequency $\omega_p\approx\omega_{eg}$ is close to the spectrally isolated $|g\rangle-|e\rangle$ transition. Nevertheless, Rabi couplings for all the dipole allowed $\{|g\rangle,|g'\rangle\}\leftrightarrow \{|e\rangle,|e'\rangle\}$ transitions are induced. Typical transient atomic populations during the $\tau_p$ probe time are represented by black dots on each level lines. }
     \label{Fig:multi} 
\end{figure}

In the following we show that for spectrally isolated $|g\rangle-|e\rangle$ transition as those in Fig.~\ref{Fig1}a and Fig.~\ref{Fig:multi} here, with $|\omega_{e'g'}-\omega_{eg}|,~|\omega_{e e'}|\gg \Gamma+\Gamma'$,  $|\omega_{g g'}|\gg \Gamma$, then the spectral phase relation by Eq.~(\ref{eq:tanbeta}) holds very precisely for smooth and long probe pulses with $\tau_p\gg 1/\Gamma$, as long as any off-resonant $|g\rangle-|e'\rangle$, $|g'\rangle-|e\rangle$ transitions only influence the atomic optical response perturbatively. Our derivations below are supported by numerical simulation of optical Bloch equations for resonant D-line interactions~\cite{Sievers2015, Qiu2022}. The results~\cite{foot:XH} for $\tau_p=1~\mu$s pulses suggest the Eq.~(\ref{eq:tanbeta}) relation supports $\overline{L}_{\rm min}< 10^{-4}$ (to compare with $\overline{L}_{\rm min}(z_A)$ in Table~(\ref{Table:cost})) for the open D1 transition of $^{39}$K in this work, improves further for resolved D2 cooling transition lines such as those for $^{87}$Rb~\cite{Wang2022b}, and becomes nearly perfect for 2-level atoms with a similar linewidth $\Gamma$. These are for weak pulses, as well as for intense pulses that strongly saturate the $|g\rangle-|e\rangle$ transition.


As illustrated in Fig.~\ref{Fig:multi},  we generally consider the interaction between a probe pulse $E_p$ and a multi-level atom. The probe frequency $\omega$ is nearly resonant to the $|g\rangle-|e\rangle$ transition with detuning $\Delta=\omega-\omega_{eg}$. In addition to the $\Omega_{eg}$ coupling, off-resonant $\Omega_{eg'}$, $\Omega_{e'g}$ couplings, as well as separate $\Omega_{e'g'}$ couplings are all induced. All the couplings contribute to the induced dipole moment $\langle {\bf d} \rangle$ and therefore the atomic polarizability $\alpha(\omega)$ to attenuate the intensity and shift the phase of $E_p$. Here, with the small detuning $\Delta$ of order $\Gamma$ and further with the aforementioned spectral isolation condition, the atomic response to $E_p$ is dominantly due to the $\rho_{eg}$ 2-level coherence. Writing the density matrix element in the frame co-rotating with the probe field, we have the single-frequency dipole moment given by Eq.~(\ref{eq:d}) in the main text. The equation of motion for $\rho_{eg}$ is according to the master equation formalism~\cite{ScullyBook}
\begin{equation}
 \begin{aligned}
       i\dot{\rho}_{eg}&=(-\Delta -i\Gamma/2)\rho_{eg}-\frac{1}{2}\Omega_{eg}(\rho_{gg}-\rho_{ee})-\\
    &~~~\sum_{g',e'}(\frac{1}{2}\Omega_{eg'}\rho_{g'g}-\frac{1}{2} \rho_{e e'} \Omega_{e'g}).
    \end{aligned}\label{eq:rhoeg}
\end{equation}
Here the Rabi frequency is defined as $\Omega_{eg}=\frac{E_p\cdot {\bf d}_{eg}}{\hbar}$, with other transitions follow similarly. Importantly, the $\Omega_{e'g}$ and $\Omega_{e g'}$ couplings in the 2nd line of Eq.~(\ref{eq:rhoeg}) are off-resonant. For ``not too strong'' pulses so that the level splittings still dominate the time scales, the dynamics of these coherences largely follow $\rho_{eg}$ and can thus be adiabatically eliminated:

\begin{equation}
\begin{aligned}
            \rho_{g'g}&\approx \frac{\Omega_{g'e}}{2\omega_{g'g}}\rho_{eg},\\
        \rho_{ee'}&\approx \frac{\Omega_{ge'}}{2\omega_{ee'}+i\Gamma+i\Gamma'}\rho_{eg}
    \end{aligned}
\end{equation}
We therefore rewrite Eq.~(\ref{eq:rhoeg}) as
\begin{equation}
    i\dot{\rho}_{eg}\approx (-(\Delta+\delta) -i(\Gamma+\gamma)/2)\rho_{eg}-\frac{1}{2}\Omega_{eg}(\rho_{gg}-\rho_{ee}),\label{eq:rhoegp}
\end{equation}
with 

\begin{equation}
    \begin{array}{l}
\delta=\sum_{g',e'}\frac{|\Omega_{e'g}|^2}{4\omega_{e e'}}-\frac{|\Omega_{e g'}|^2}{4\omega_{g'g}}\\
\gamma=\sum_{e'}\frac{|\Omega_{e'g}|^2}{4\omega_{e e'}^2}\Gamma'.
    \end{array}
\end{equation}
Equation~(\ref{eq:rhoegp}) is coupled to $\rho_{gg}(t),\rho_{ee}(t)$ dynamics that we have not written out explicitly. However, for understanding $\beta_0={\rm arg}(\alpha)$, it suffices to integrate Eq.~(\ref{eq:rhoegp}) to have
\begin{equation}
    \rho_{eg}(t)\approx \frac{i}{2}\int_0^t \Omega_{eg}(\tau)(\rho_{gg}(\tau)-\rho_{ee}(\tau)) e^{- i\tilde \Delta (t-\tau)}{\rm d}\tau.\label{eq:rhot}
\end{equation}
Here $\tilde \Delta=(\Delta+\delta) + i(\Gamma+\gamma)/2$. We then evaluate the nominator of Eq.~(\ref{eq:alpha}) in the main text, 
\begin{equation}
    \begin{aligned}
\alpha &\propto \int_0^{\tau_p}{\rm d} t E^*_p(t)\cdot {\bf d}_{g e}\rho_{eg}(t)\\
&\approx \frac{i\hbar }{2}\int_0^{\tau_p}{\rm d} t |\Omega_{eg}(t)| \int_0^t \Lambda(\tau) e^{- i\tilde \Delta (t-\tau)}{\rm d}\tau.
    \end{aligned} \label{eq:alpha2}
\end{equation}
Here $\Lambda(\tau)=|\Omega_{eg}(\tau)|(\rho_{gg}(\tau)-\rho_{ee}(\tau))$. 

To arrive at Eqs.~(\ref{eq:rhot})(\ref{eq:alpha2}) with smooth $E_p(t)$, we have ignored the time-dependence of $\delta(t),\gamma(t)$ so that $\dot{\tilde \Delta}= 0$. Accordingly, we regard $\delta,\gamma$ as average values for Eq.~(\ref{eq:tanbeta}). Furthermore, assuming smooth pulse with $\Lambda^{(n)}(0)=0$, Eq.~(\ref{eq:alpha2}) can be evaluated with integration by parts,
\begin{equation}
\begin{aligned}
\alpha &\propto i\int_0^{\tau_p} {\rm d} t |\Omega_{eg}(t)| \left[\frac{\Lambda(t)}{-i \tilde \Delta}+
\int_0^t \frac{\Lambda'(\tau)}{i \tilde \Delta } e^{- i\tilde \Delta (t-\tau)}{\rm d}\tau\right] \\
&=i\int_0^{\tau_p} {\rm d} t |\Omega_{eg}(t)| \left[\frac{\Lambda(t)}{-i \tilde \Delta}+
\frac{\Lambda'(t)}{ \tilde \Delta^2}-
 \int_0^t \frac{\Lambda''(\tau)}{\tilde \Delta^2 } e^{- i\tilde \Delta (t-\tau)}{\rm d}\tau\right].
 \end{aligned}\label{eq:alpha3}
\end{equation}
For the smooth pulse with $\tau_p\gg 1/\Gamma$, then the leading term dominates in the integration.  Assuming $\rho_{gg}>\rho_{ee}$, $\alpha$ has a phase angle $\beta_0=-{\rm arg}(1/\tilde \Delta)$ that obeys Eq.~(\ref{eq:tanbeta}). Here, it is interesting to note that for symmetric $|\Omega_{eg}(t)|\approx |\Omega_{eg}(\tau_p-t)|$ pulse profiles, the second term $\Omega_{eg}(t) \frac{\Lambda'(t)}{\tilde \Delta^2}$ tends to average itself out with the integration. The average effect works for both weak and strong, off and resonant pulses, as long as $\rho_{gg}(t)-\rho_{ee}(t)$ approximately follows the same symmetry at the $\tau_p$ scale. Clearly, this average effect works best for closed transitions, and is compromised for open transitions as in this work, due to the population quenching (atoms leaving $\{|g\rangle, |e\rangle\}$ to other states at the $1/\Gamma$ time scale). 

We finally remark that the Eq.~(\ref{eq:d}) approximation requires the $|g\rangle-|e\rangle$ transition is strong, with large enough dipole element ${\bf d}_{eg}$ comparing to nearby off-resonant transitions. For atomic states with Zeeman degeneracy, the  multi-level interaction with $E_p$ can be decomposed into degenerate 2-level dynamics~\cite{Shore2014} to ``monomorphously'' contribute to 
the optical response. The Eq.~(\ref{eq:tanbeta}) relation is therefore still valid. Finally, the Eq.~(\ref{eq:tanbeta}) relation is robust against inhomogeneous broadening effects, such as those due to Zeeman shifts, Doppler shifts, or a finite $E_p$ laser linewidth (as in this work~\cite{He2020a}), as long as these broadenings are small comparing to the transition linewidth $\Gamma$. These conclusions are supported by numerical simulations~\cite{Sievers2015, Qiu2022}, with details to be presented in a separate paper~\cite{foot:XH}.


\section{Simulation of complex phase shift by 2-level samples}
\label{sec:simu}
In this section we provide details on the simulation of $E_p$ propagation through a weakly-driven 2-level atomic sample. In all the simulations leading to Fig.~\ref{Fig1}c, Fig.~\ref{Fig:spec}(a-c), the atomic sample assumes a Gaussian distribution $\rho({\bf r})$ with a spatial width matching the experimental situation for the $^{39}$K atomic samples (Sec.~\ref{sec:exp}). The numerical propagation assumes a local complex index $n({\bf r})=\sqrt{1+\varrho({\bf r})  \alpha}$ with atomic polarizability:
\begin{equation}
    \alpha=\frac{3\lambda_p^3}{8\pi^2}\frac{(-\Delta+i\frac{\Gamma}{2})\frac{\Gamma}{2}}{\Delta^2+(1+s)\frac{\Gamma^2}{4}}
\end{equation}
Here an $s-$parameter is introduced to effectively account for the saturation effect in the simulation~\cite{ScullyBook}, with $s=0.5$ for the Fig.~\ref{Fig:spec}(d-f) simulation.

Taking $E_p(z)$ from experimental pre-characterization (Sec.~\ref{sec:rec}), numerical propagation of $E_p(z)$ across the sample follows a split-operator method~\cite{Blanes2000}, by interleaving the free-space propagation (Eq.~(\ref{eq:ASM}))  $E_{\rm tot}(z+dz)=U(dz)E_{\rm tot}(z)$ with spatial-dependent phase shift  $E_{\rm tot}(z+{\rm d}z)=E_{\rm tot}(z)e^{i (n(z)-1) k_p {\rm d}z}$ in small steps ${\rm d}z$. Starting from $E_{\rm tot}=E_p(z_A-l_z)$, we obtain $E_{\rm tot}=E_p(z_A+l_z)$ across the full sample length in the model. The full field is then decomposed into $E_{\rm tot}=E_p+E_s$ to find the ``true'' complex phase shift $\varphi(z_A)$ at the atomic central plane $z=z_A$ according to Eq.~(\ref{eq:varphi}), as well as to numerically generate $I_{1,2}$ at the camera plane $z=z_H$ for verifying our holographic reconstruction and sample plane localization algorithms.

\section{Experimental details}
\label{sec:expDetail}

\begin{figure}[htbp]
    \centering
    \includegraphics[width=.75\linewidth]{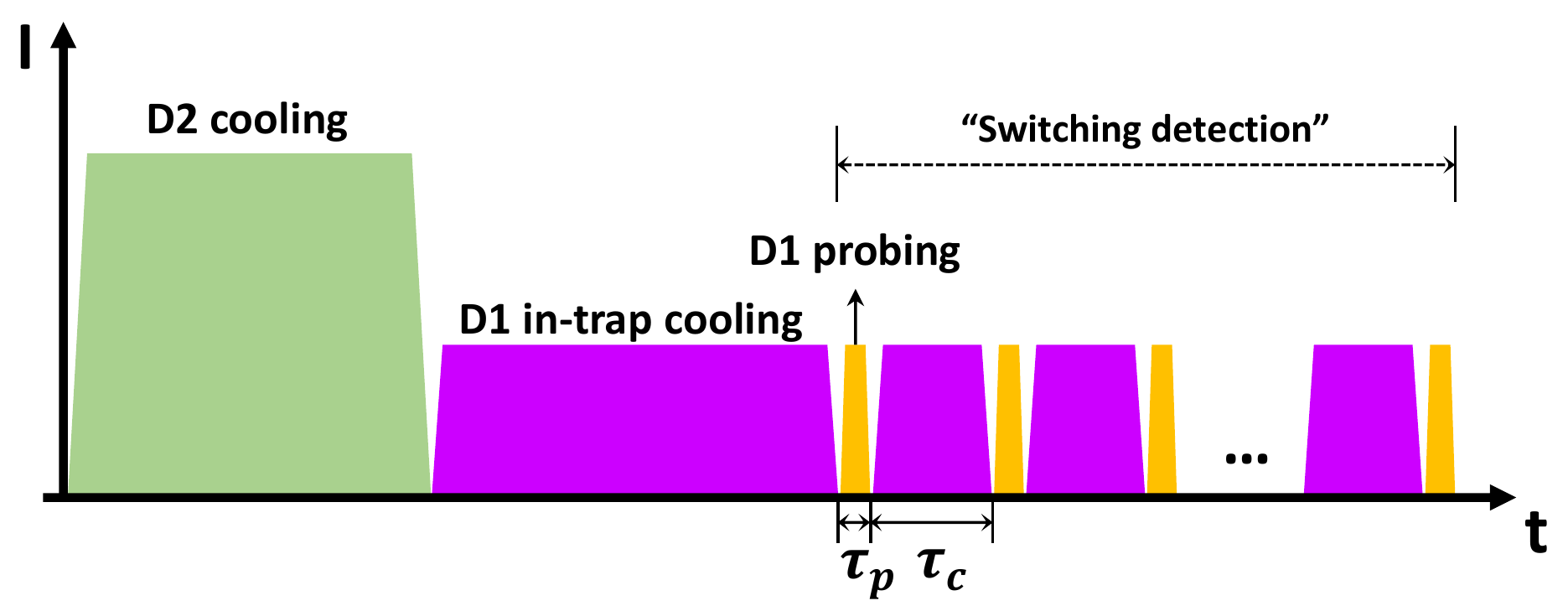}
    \caption{The timing sequence of a single experimental measurement cycle, including D2 cooling and trapping, D1 in-trap cooling and probing by "switching detection" (axes not to scale). During the "switching detection", the D1 probing and D1 in-trap cooling are interleaved with $\tau_p=1~\mu$s and $\tau_c=4~\mu$s respectively.}
     \label{Fig:Timing} 
\end{figure}

\begin{figure}[htbp]
    \centering
    \includegraphics[width=.75\linewidth]{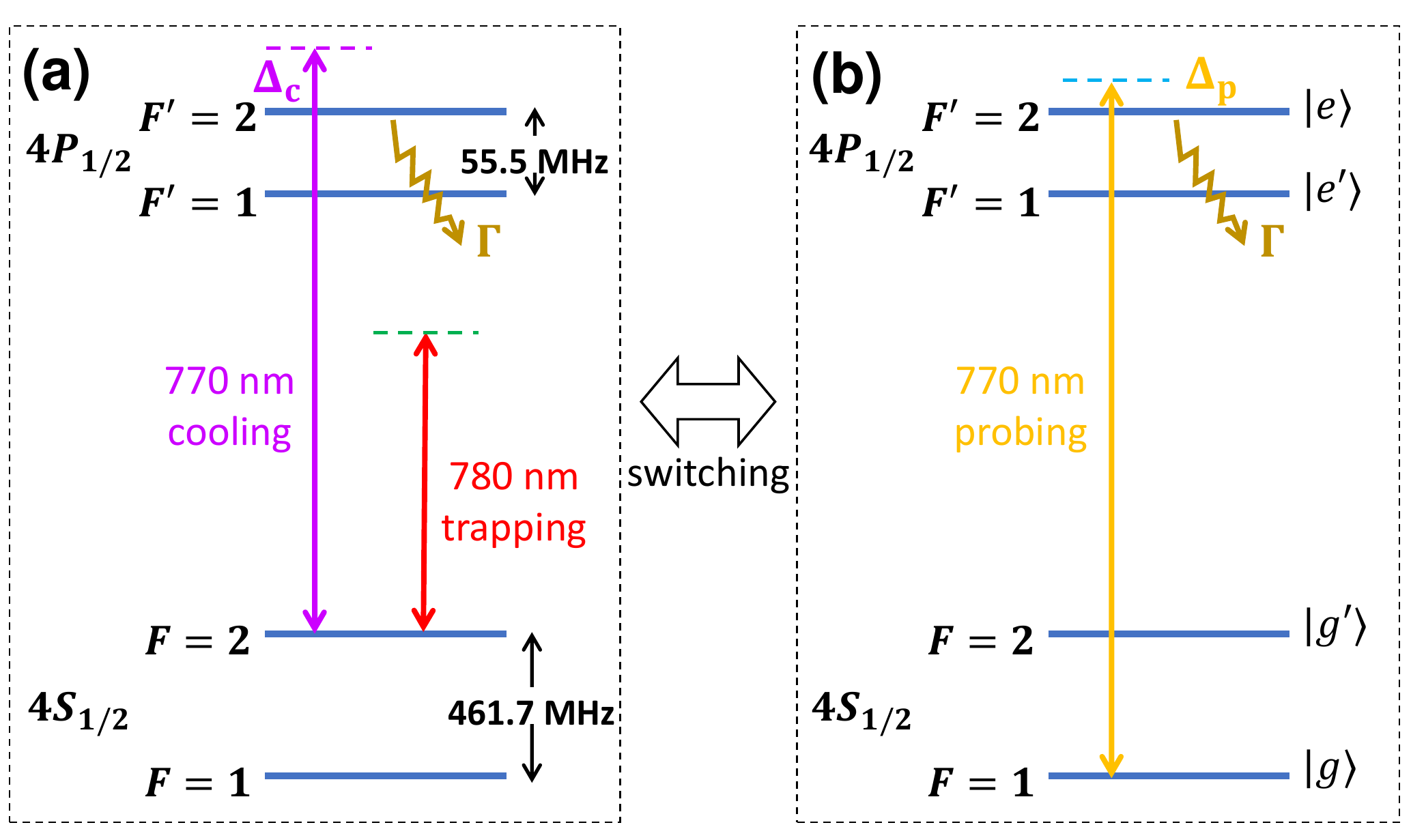}
    \caption{Level diagram for the D1 "switching detection" at the $^{39}$K D1 line in this work. (a): D1 in trap cooling. (b): Probing. The atomic states in (b) are labeled in according to Fig.~\ref{Fig1}b in the main text.} 
     \label{Fig:D1switch} 
\end{figure}

In this section we provide additional experimental details for holographic imaging of $^{39}$K atoms. As mentioned in Sec.~\ref{sec:exp}, the $\{I_{1,2}^{(j)}\}$ data at detunings $\{\Delta_j\}$ are obtained with repeated cycles of atomic sample preparation and measurements. A single cycle takes about 2 seconds time. The timing sequence for a single cycle is shown in Fig.~\ref{Fig:Timing}. Firstly, a standard magneto-optical trap (MOT)~\cite{MetcalfBook} with D2 line cooling captures $\sim 10^6$ atoms with a temperature $T\approx $ 0.5~mK. We then apply D1 molasses cooling~\cite{Sievers2015}, with $\sim 100~$mW of 770~nm light delivered to the atoms with the same MOT beams. During this cooling stage, the D1 molasses contain two sidebands to address both hyperfine ground states $4S_{1/2}, F=1,2$ at Raman 2-photon resonance~\cite{Sievers2015}. The single-photon detuning to $F'=2$ is set as $\Delta_c=2\pi\times16$~MHz.  To create the microscopic sample for the imaging study, an optical dipole trap (ODT) at wavelength 780~nm is switched on simultaneously. The ODT is fairly strongly focused by another NA=0.3 lens array to reach a Gaussian waist of $\approx$1.5 $\mu$m at the  MOT center, with a trap depth $\approx$ 15 MHz.  This combined D1 in-trap cooling of about 3~ms is able to prepare the $N=10^3$ atomic sample at a temperature of $T\approx$ 10 $\mu$K. Next, in the holographic imaging step, we perform "switching detection" by interleaving $\tau_p=1~\mu$s D1 probe pulses with $\tau_c=4~\mu$s D1 single-sideband cooling + 780 trapping pulse, for a total exposure time of 1 ms. The level diagrams for the light-atom interaction during probing and cooling are summarized in Fig.\ref{Fig:D1switch}. In particular, during the D1 in-trap cooling that only addresses the $F=2$ hyperfine level, atoms are not only cooled and trapped, but also ``depumped'' back to $F=1$ for the next $\tau_p$ probe.









\bibliography{References}



\end{document}